\documentclass[a4paper,11pt]{article}
\usepackage{amssymb,amsmath,amsthm,amsfonts}
\usepackage{bookmark}
\usepackage{mathrsfs}
\usepackage{multirow}
\usepackage{graphicx}
\usepackage{authblk}
\usepackage{indentfirst}
\usepackage{multicol}
\usepackage{tabu}
\usepackage{url}
\usepackage{fancyhdr}
\usepackage[numbers]{natbib}
\usepackage[all]{xy}
\usepackage{mathtools}
\usepackage[english]{babel}
\usepackage{colortbl}
\usepackage{caption}
\usepackage{hyperref}\usepackage{subcaption}
\usepackage{tikz}\usepackage{float}

\usepackage{amsthm} %

\newtheorem{example}{Example}[section]

\theoremstyle{plain} %

\newtheorem{definition}{Definition}[section]

\newcommand{\im}{{\mathrm{im}\hspace{0.1em}}}

\usepackage{xr}
\makeatletter
    \newcommand*{\addFileDependency}[1]{
    \typeout{(#1)}
    \@addtofilelist{#1}
    \IfFileExists{#1}{}{\typeout{No file #1.}}
    }
\makeatother

\title{Topological sequence analysis of genomes: category theory approaches}
\author[1,2]{Jian Liu \thanks{This author's work was completed at Michigan State University}}
\author[2]{Li Shen}
\author[2]{Mushal Zia}
\author[2,3,4]{Guo-Wei Wei \thanks{Corresponding author: weig@msu.edu}}
\affil[1]{Mathematical Science Research Center, Chongqing University of Technology, Chongqing 400054, China}
\affil[2]{Department of Mathematics, Michigan State University, MI 48824, USA}
\affil[3]{Department of Electrical and Computer Engineering, Michigan State University, MI 48824, USA}
\affil[4]{Department of Biochemistry and Molecular Biology, Michigan State University, MI 48824, USA}

\makeatletter
    \renewcommand*{\@fnsymbol}[1]{\ensuremath{\ifcase#1\or \dagger\or *\or *\or
   \mathsection\or \else\@ctrerr\fi}}
\makeatother
\date{}

\begin{document}
    \maketitle
  \paragraph{Abstract}

Sequence data, such as DNA, RNA, and protein sequences, exhibit intricate, multi-scale structures that pose significant challenges for conventional analysis methods, particularly those relying on alignment or purely statistical representations. In this work, we introduce category-based topological sequence analysis (CTSA ) of genomes. CTSA  models a sequence as a resolution category, capturing its hierarchical structure through a categorical construction. Substructure complexes are then derived from this categorical representation, and their persistent homology is computed to extract multi-scale topological features. Our models depart from traditional alignment-free approaches by incorporating structured mathematical formalisms rooted in sequence topology. The resulting topological signatures provide informative representations across a variety of tasks, including the phylogenetic analysis of  SARS-CoV-2 variants and the prediction of protein-nucleic acid binding affinities. Comparative studies were carried out against six state-of-the-art methods. Experimental results  demonstrate that CTSA  achieves excellent and consistent performance in these tasks, suggesting its general applicability and robustness. Beyond sequence analysis, the proposed framework opens new directions for the integration of categorical and homological theories for biological sequence analysis.

    \paragraph{Keywords}
     Topological sequence analysis, category, substructure complex, alignment-free, DNA sequence.

\footnotetext[1]
{ {\bf 2020 Mathematics Subject Classification.}  	Primary  55N31; Secondary 62R40, 68T09.
}

    \newpage
    \tableofcontents
    \newpage

\section{Introduction}\label{section:introduction}

Biological sequences, such as DNA, RNA, and proteins, lie at the heart of molecular biology and govern the functional behavior of cells and organisms. These sequences are not merely strings of symbols; they encode complex hierarchical and structural information that shapes biological functions such as gene regulation, protein binding, and enzymatic activity. For instance, the binding affinity between a protein and a nucleic acid sequence often depends not only on specific motifs but also on broader contextual patterns, repetitions, and structural relationships within the sequence. Traditional methods that rely solely on alignment or motif detection may fail to capture these intricate, multiscale properties.

To analyze biological sequences, a variety of traditional computational methods have been developed. Alignment-based methods, such as BLAST~\cite{ye2006blast} and clustal~\cite{larkin2007clustal}, are widely used for sequence comparison and homology detection. They excel at identifying conserved regions and are highly effective for sequences with strong evolutionary relationships. However, these methods rely on global or local alignments and often struggle with highly divergent or structurally rearranged sequences. In contrast, alignment-free methods, such as natural vector method~\cite{sun2021geometric}, Markov models~\cite{qi2004whole,reinert2009alignment}, and feature frequency profile-based approaches~\cite{jun2010whole,yu2013real,sims2009alignment}, offer scalable and flexible alternatives that bypass the need for pairwise alignment. These methods are well-suited for large-scale comparative genomics, classification of viral genomes, and studies involving highly diverse or recombined sequences. Nonetheless, most alignment-free approaches emphasize content-level features and may overlook the positional or relational structure among subsequences, leading to low accuracy.

Topological Data Analysis (TDA) is a class of techniques rooted in algebraic topology, designed to extract and analyze the intrinsic topological features of complex data~\cite{carlsson2009topology,edelsbrunner2010computational,zomorodian2004computing}. Traditionally, TDA has been extensively applied to point cloud data, enabling the study of shapes, clusters, and voids at multiple scales.
Recently, persistent spectral theory, also known as persistent Laplacian, has been introduced \cite{wang2020persistent, wei2025persistent}. This approach has been successfully applied to the analysis of  SARS-CoV-2 variants \cite{wei2023persistent2}.
More recently, the scope of TDA has also been expanded to encompass to  persistent magnitude \cite{bi2024persistent}, persistent Mayer homology \cite{shen2024persistent,feng2025mayer}, and knot-based data analysis, where topological invariants such as knots and tangles serve as informative structures for understanding complex entanglements in physical, biological, or abstract systems~\cite{liu2024persistent,shen2024knot,shen2024evolutionary}.

Sequence data, particularly in biological contexts such as genomic or proteomic sequences, often exhibit intricate structural complexity that poses challenges for conventional analytical methods. Despite the growing success of TDA in capturing global features of complex data, its application to sequence data remains relatively underexplored. Recently, Hozumi and Wei introduced a topological sequence analysis (TSA) approach via persistent homology and persistent Laplacian to reveal the structure of the genome space ~\cite{hozumi2024revealing}.

In this work, we aim to develop a fundamental TSA framework using category theory. Our goal is to construct models and methods that can effectively capture and analyze the underlying topological features of sequences, thereby opening new categorical perspectives for structural understanding and data-driven discovery. To bridge the strengths of both alignment-based and alignment-free paradigms, we propose a novel framework called category-based topological sequence analysis (CTSA). CTSA  is an alignment-free but retains the ability to capture fine-grained positional and structural relationships typically accessible only through alignment. At the core of our approach is the observation that many biologically patterns arise from relationships between k-mers of varying lengths. Rather than fixing a single k-mer size, we analyze the entire hierarchy of k-mer spaces across different resolutions. We formalize this hierarchy using a mathematical categorification: contiguous subsequences and their positions in the parent sequence are modeled as objects in a category, and inclusion-respecting relationships are modeled as morphisms. This construction, called the resolution category, encodes both the content and the relative position of each subsequence within the sequence.

From this categorical foundation, we derive a family of substructure complexes that realize the resolution category as topological objects. Equipped with natural distance functions on objects in the category reflecting similarity in structure, position, and frequency, we construct filtered simplicial complexes whose evolving shapes capture multiscale structural information in the sequence. Persistent homology is then applied to these complexes to extract robust, quantitative topological features. The resulting sequence representation retains the scalability and flexibility of alignment-free methods while incorporating positional and compositional information typically preserved in alignment-based approaches.

We validate the effectiveness of CTSA in two representative biological tasks. In the prediction of protein-nucleic acid binding affinities, we integrate CTSA features from DNA or RNA sequences with ESM2 embeddings~\cite{rives2021biological,lin2022language} of protein sequences, and use the combined representations in supervised learning models. Our method achieves the state-of-the-art predictive accuracy, with Pearson correlation of 0.709 and RMSE of 1.29 kcal/mol. These results are better than those  of existing baselines, including methods using DNABERT~\cite{ji2021dnabert}, a pretrained transformer model for DNA sequences, as well as other alignment-free approaches based on k-mer topology \cite{hozumi2024revealing}. In the second task, the  clustering of SARS-CoV-2 genomic variants, CTSA alone is able to separate known variant clades with 100\% accuracy, outperforming five state-of-the-art alignment-free methods. These results demonstrate that CTSA not only scales to long and complex sequences, but also effectively preserves sequence-level structural patterns that are critical for downstream biological analysis, making it a powerful tool for both predictive and comparative genomics.

\section{Preliminaries}

The central idea of this work is to apply categorical and topological methods to the study of sequences, which are especially prevalent in biological contexts. In this section, we provide a brief introduction to some fundamental concepts and ideas from category theory and persistent homology that underpin our approach.

\subsection{Category theory}

Category theory provides a systematic mathematical framework for describing the objects within a complex system and the relationships between them. Roughly speaking, a category consists of a collection of objects together with morphisms between them, subject to certain axioms.

Mathematically, a category $\mathcal{C}$ consists of the following data:
\begin{itemize}
    \item A class of objects, denoted $\mathrm{Ob}(\mathcal{C})$;
    \item For each pair of objects $X, Y \in \mathrm{Ob}(\mathcal{C})$, a set of morphisms $\mathrm{Hom}_{\mathcal{C}}(X, Y)$;
    \item For each object $X$, an identity morphism $\mathrm{id}_X \in \mathrm{Hom}_{\mathcal{C}}(X, X)$;
    \item A composition law: for morphisms $f \in \mathrm{Hom}_{\mathcal{C}}(X, Y)$ and $g \in \mathrm{Hom}_{\mathcal{C}}(Y, Z)$, a composite morphism $g \circ f \in \mathrm{Hom}_{\mathcal{C}}(X, Z)$.
\end{itemize}
These data satisfy the following axioms:
\begin{enumerate}
    \item[(i)] Associativity: For any composable morphisms $f \colon X \to Y$, $g \colon Y \to Z$, and $h \colon Z \to W$, we have
    \begin{equation}
    h \circ (g \circ f) = (h \circ g) \circ f;
    \end{equation}
    \item[(ii)] Identity: For any morphism $f \colon X \to Y$, we have
    \begin{equation}
    \mathrm{id}_Y \circ f = f = f \circ \mathrm{id}_X.
    \end{equation}
\end{enumerate}

While the formal definition of a category is abstract, the underlying ideas have concrete and intuitive interpretations in a wide range of applied settings. At its core, category theory is about understanding not only the nature of individual components in a system, but also how these components are connected and interact with one another through well-defined relationships.

In practical terms, objects in a category can represent data types, computational states, geometric shapes, physical systems, or even conceptual entities. Morphisms then encode meaningful transformations, mappings, or processes between these objects. For example, in computer science, one might consider objects as data structures and morphisms as algorithms converting one structure into another. In physics, objects could represent spaces or states, with morphisms modeling symmetries or time evolution. In network theory, nodes can be treated as objects and edges as morphisms capturing interactions or flows.

This abstraction is powerful because it allows us to compare, compose, and reason about systems in a unified way, regardless of the specific domain. Importantly, category theory emphasizes the structure-preserving nature of transformations, making it particularly suitable for studying complex systems where coherence, compatibility, and compositionality are essential.

In this work, we use categories to describe structured sequences, and we further construct associated \emph{substructure complexes}, which encode geometric and topological information embedded within the sequences. These constructions form the bridge between categorical representations and the computation of multiscale topological features.

\subsection{Persistent Homology}

Persistent homology, developed over the past two decades, has become one of the most classical tools in  TDA. In this section, we briefly review the fundamental concepts and ideas underlying persistent homology.

Homology is a fundamental topological invariant that captures intrinsic structural features of a space, such as connected components, loops, and voids. These features are invariant under continuous deformations, meaning they are independent of the choice of coordinate system and remain unchanged under stretching or bending. As a result, homology provides a robust description of the global structure of data, highlighting stable and meaningful features.

Persistent homology extends this idea to multiscale data analysis. Given a dataset and a scale parameter $\epsilon$, one constructs a family of topological spaces $\{X_\epsilon\}_{\epsilon \geq 0}$ that reflect the connectivity of the data at different scales. As $\epsilon$ increases, new topological features (homology generators) may appear and existing ones may vanish. Each such feature is characterized by a birth time $r$ and a death time $s$, indicating the scale at which it appears and disappears. The collection of these intervals, known as the persistence diagram or barcode, encodes the topological evolution of the data across scales.

More precisely, let $X$ be a topological space, typically a simplicial complex in applications. Let $f \colon X \to \mathbb{R}$ be a real-valued function defined on $X$, often representing some measurement or filtration. One considers the sublevel set filtration $\{X_a := f^{-1}((-\infty, a])\}_{a \in \mathbb{R}}$, which yields a nested sequence of spaces. The inclusion maps $X_a \hookrightarrow X_b$ for $a \leq b$ induce homomorphisms on homology groups
\begin{equation}
H_k(X_a) \longrightarrow H_k(X_b)
\end{equation}
for each dimension $k \geq 0$. These maps track the evolution of $k$-dimensional topological features (e.g., connected components, loops, voids) as the scale parameter $a$ increases.

\begin{definition}
Let $\{X_a\}_{a \in \mathbb{R}}$ be a filtration of a topological space $X$. The \emph{$k$-th persistent homology} of this filtration is the family of images
\begin{equation}
H_k^{a,b} := \mathrm{Im}\left(H_k(X_a) \to H_k(X_b)\right),
\end{equation}
where $a \leq b$ and the map is induced by the inclusion $X_a \hookrightarrow X_b$.
\end{definition}
Each class in $H_k^{a,b}$ is a $k$-dimensional homology class that is born at or before time $a$ and still exists at time $b$. The collection of all such intervals $[a, b)$, corresponding to the lifespans of homology classes, defines the \emph{barcode} or \emph{persistence diagram}.

In practice, persistent homology is often computed over a discrete filtration such as a sequence of simplicial complexes (e.g., Vietoris-Rips or \v{C}ech complexes), and with coefficients in a field (typically $\mathbb{F}_2$). Classical persistent homology is commonly computed using libraries such as \texttt{GUDHI} and \texttt{Ripser}. However, the implementation of more recent or advanced methods often requires extending or modifying these existing frameworks.

\section{The resolution of sequences}

In this section, we introduce the resolution category of a sequence. This category provides an algebraic framework that captures the sequence's multi-scale structure and allows for its reconstruction. We also define various distance measures between objects in the resolution category. These distances reveal the multi-scale structure of sequences and establish a link to topological analysis. They lay the groundwork for developing a theory of persistent homology on sequences.

\subsection{The resolution category}\label{section:resolution}
Let $X$ be a non-empty finite set. For each $n \geq 0$, denote by $EX_n$ the set of all $(n+1)$-tuples $(x_0, x_1, \dots, x_n)$ with each $x_i \in X$. Define
\begin{equation}
EX_{\leq n} := \coprod_{k=0}^{n} EX_k,
\end{equation}
which represents the disjoint union of all sequences of elements in $X$ of length at most $n+1$.

Given a finite sequence $\xi = (x_0, x_1, \dots, x_m)$ with $x_i \in X$, we are interested in capturing all of its contiguous subsequences (i.e., substrings), along with their positions of occurrence in $\xi$. For each such contiguous subsequence $\sigma$ that appears in $\xi$, and for each position $i$ such that $\xi[i : i + |\sigma|] = \sigma$, we associate the pair $(\sigma, i)$, which we refer to as the anchored subsequence $\sigma$ at position $i$.

We now define a category $\mathcal{C}_\xi$, referred to as the \emph{resolution category} of $\xi$, whose objects are all such indexed contiguous subsequences.

An object of $\mathcal{C}_\xi$ is a pair $(\sigma, i)$, where $\sigma = (x_{i}, x_{i+1}, \dots, x_{i+\ell})$ is a contiguous subsequence of $\xi$, and $i$ is its starting index in $\xi$. A morphism
\begin{equation}
(\sigma, i) \longrightarrow (\tau, j)
\end{equation}
in $\mathcal{C}_\xi$ exists if and only if $\sigma$ is a contiguous subsequence of $\tau$, and the occurrence of $\sigma$ within $\tau$ is consistent with their respective positions in $\xi$. That is, there exists $0 \leq k \leq |\tau| - |\sigma|$ such that
\begin{equation}
\sigma = (x_{j + k}, x_{j + k + 1}, \dots, x_{j + k + |\sigma|-1})
\quad \text{and} \quad i = j + k.
\end{equation}
In this case, we declare a unique morphism from $(\sigma, i)$ to $(\tau, j)$.

From this construction, it is evident that the category $\mathcal{C}_{\xi}$ possesses a hierarchical structure determined by the lengths of the subsequences. Let the collection of objects of length $n+1$ be denoted by $\mathcal{S}_{\xi}(n)$; then the set of all objects in $\mathcal{C}_\xi$ is given by
\begin{equation}
\mathrm{ob}(\mathcal{C}_{\xi}) = \coprod_{k \geq 0} \mathcal{S}_{\xi}(k).
\end{equation}
For any $k \leq m$, the hierarchy $\mathcal{S}_{\xi}(k)$ contains all indexed subsequences of length $k+1$ and can collectively recover the original sequence $\xi$. For instance, the lowest level $\mathcal{S}_{\xi}(0)$ consists of all elements of $\xi$ indexed by their positions and thereby suffices to reconstruct $\xi$. Thus, we interpret the category $\mathcal{C}_{\xi}$ as a positional resolution of the sequence $\xi$.

It is straightforward to verify that $\mathcal{C}_\xi$ is a small category. Each object admits an identity morphism, and morphism composition corresponds to subsequence inclusion respecting positions.

\begin{example}
Let $X = \{a, b, c\}$ be a finite set, and consider the sequence $\xi = (a, b, c, b, a)$. For convenience, we write this sequence as the word \texttt{abcba}. We construct the resolution category $\mathcal{C}_\xi$ associated with $\xi$.

The objects of $\mathcal{C}_\xi$ are pairs $(\sigma, i)$, where $\sigma$ is a contiguous subword of $\xi$, and $i$ is the starting position of this subword in $\xi$ (positions start from $0$). For example,
\begin{itemize}
    \item $(\texttt{c}, 2)$ denotes \texttt{c} at position $2$;
    \item $(\texttt{bc}, 1)$ denotes the subword \texttt{bc} starting at position $1$;
    \item $(\texttt{abc}, 0)$ denotes \texttt{abc} at position $0$.
\end{itemize}

A morphism $(\sigma, i) \to (\tau, j)$ exists in $\mathcal{C}_\xi$ if and only if $\sigma$ is a contiguous subword of $\tau$, and the alignment of positions within $\xi$ is respected (i.e., $\sigma$ begins at position $i = j + k$ inside $\tau$). For example:
\begin{itemize}
    \item $(\texttt{b}, 1) \to (\texttt{bc}, 1)$, since \texttt{b} is a prefix of \texttt{bc} starting at position $1$;
    \item $(\texttt{cb}, 2) \to (\texttt{bcb}, 1)$, since \texttt{cb} is a suffix of \texttt{bcb} starting at $2$;
    \item $(\texttt{a}, 0) \to (\texttt{abc}, 0)$, since \texttt{a} is a prefix of \texttt{abc} starting at $0$.
\end{itemize}

The morphism structure among these indexed subwords can be visualized with a diagram resembling a poset:
\begin{equation}
\xymatrix@C=3em{
\texttt{a} \ (0) \ar@{->}[d] &\texttt{b} \ (1)\ar@{->}[d]\ar@{->}[dl]   & \texttt{c}\ (2)\ar@{->}[d]\ar@{->}[dl] & \texttt{b} \ (3) \ar@{->}[d]\ar@{->}[dl]& \texttt{a} \ (4)\ar@{->}[dl]\\
\texttt{ab} \ (0) \ar@{->}[d] & \texttt{bc} \ (1) \ar@{->}[d]\ar@{->}[dl]   & \texttt{cb} \ (2) \ar@{->}[d]\ar@{->}[dl]& \texttt{ba} \ (3) \ar@{->}[dl]\\
\texttt{abc} \ (0)&  \texttt{bcb} \ (1) & \texttt{cba} \ (2) &  &
}
\end{equation}
\textcolor[rgb]{1.00,0.00,0.00}The above diagram shows morphisms between objects of word length $0$, $1$, and $2$, where objects of length $0$ map to those of length $1$, and objects of length $1$ map to those of length $2$. In addition, compositions of morphisms are also morphisms; for example, the morphism $(\texttt{b},3) \to (\texttt{bcb},1)$ is the composition of the morphism $(\texttt{b},3) \to (\texttt{cb},2)$ and the morphism $(\texttt{cb},2) \to (\texttt{bcb},1)$.

This example illustrates how the resolution category $\mathcal{C}_\xi$ encodes both the combinatorial and positional structure of the contiguous subwords of a given sequence.
\end{example}

The category $\mathcal{C}_\xi$ encodes not only the hierarchical structure of subsequences under inclusion, but also the temporal (or positional) information of each subsequence within the ambient sequence $\xi$. This category can be used as an indexing category to construct diagrams in topological data analysis. For instance, one may associate to each object $(\sigma, i)$ a value $f(\sigma, i) \in \mathbb{R}$, such as frequency, entropy, or interaction score, and study the persistent homology of the corresponding filtered complex.

\subsection{Distances on resolution category}

In this section, we define several distances on the objects and morphisms in the resolution category. These distances are designed to quantify the similarity between subsequences, considering both their structural and positional information within the original sequence. Such distances are fundamental for constructing filtrations on complexes built over sequences, and they play a crucial role in analyzing the underlying topological structure of sequences.

To compare the dissimilarity between two sequences, common distance measures include the Levenshtein distance, the longest common subsequence (LCS) distance, and the Hamming distance~\cite{hamming1950error,hirschberg1975linear,levenshtein1966binary}.

The Levenshtein distance is a widely used distance that quantifies the minimum number of edit operations (insertions, deletions, and substitutions) required to transform one sequence into another. The LCS distance, as the name suggests, is based on the length of the longest common subsequence shared by two sequences, providing a measure of their structural similarity. Specifically, it is defined as
\begin{equation}
d_{\mathrm{LCS}}(X,Y) = 1 - \frac{\mathrm{LCS}(X,Y)}{\min(\# X, \# Y )},
\end{equation}
where \(\mathrm{LCS}(X,Y)\) denotes the length of the longest common subsequence of sequences \(X\) and \(Y\). Here, \(\# X\) denotes the length (i.e., the number of elements) of the sequence \(X\).

The Hamming distance, particularly suitable for sequences of equal length, counts the number of positions at which the corresponding elements differ. However, these classical distances primarily focus on the intrinsic differences between two sequences themselves. They are not designed to capture the differences in the occurrences or contextual roles of two sequences within a larger sequence.

More specifically, our goal is to understand how two subsequences differ in terms of their presence and behavior inside a larger sequence $\xi$, as captured by the associated resolution category. In this context, we seek a refined notion of distance that compares not only the content of two subsequences, but also how they are situated, repeated, and structurally embedded within the ambient sequence. This perspective shifts the focus from sequence-level comparison to a categorical-level comparison based on the resolution structure induced by $\xi$.

\subsubsection*{Intersection distance}

Let $(\sigma_1, i_1)$ and $(\sigma_2, i_2)$ be two objects in the resolution category $\mathcal{C}_\xi$, where $\sigma_1$ and $\sigma_2$ are contiguous subsequences of $\xi$. Consider all objects $(\sigma_0, i_0)$ admitting morphisms
\begin{equation}
(\sigma_1, i_1) \leftarrow (\sigma_0, i_0) \rightarrow (\sigma_2, i_2)
\end{equation}
in $\mathcal{C}_\xi$. Let $L = \max\limits_{(\sigma_0, i_0)} \mathrm{length}(\sigma_0)$ be the maximum length among all such intersection objects.

We define the \emph{intersection distance} between $(\sigma_1, i_1)$ and $(\sigma_2, i_2)$ as the normalized complement of this overlap
\begin{equation}
d_{\mathrm{int}}((\sigma_1, i_1), (\sigma_2, i_2)) := 1 - \frac{L}{\min\{ \mathrm{length}(\sigma_1), \mathrm{length}(\sigma_2) \}}.
\end{equation}

This distance lies in $[0,1]$, and takes the value $0$ if and only if the shorter of the two subsequences is fully contained in a common intersection object. It increases as the shared structure between the two objects becomes smaller.

\subsubsection*{Position distance}

Beyond content-based comparisons, positional information also plays a critical role in analyzing the behavior of subsequences within a larger sequence.

We define the \emph{position distance} between two objects $(\sigma_1, i_1)$ and $(\sigma_2, i_2)$ in the resolution category as
\begin{equation}
d_{\mathrm{pos}}((\sigma_1, i_1), (\sigma_2, i_2)) = |i_1 - i_2|.
\end{equation}

This distance simply measures the absolute difference in their starting positions within the sequence $\xi$. A larger value indicates that the two subsequences are located farther apart in $\xi$. The position distance is non-negative, symmetric, and satisfies the triangle inequality, thereby qualifying as a valid distance on the set of objects indexed by their positions.

\subsubsection*{Frequency distance}

Fix a contiguous subsequence $\sigma$. For a given ambient sequence $\xi$, we consider the resolution category $\mathcal{C}_\xi$. Let $S(\sigma)$ denote the set of all objects of the form $(\sigma, i)$ in $\mathcal{C}_\xi$, i.e.,
\begin{equation}
S(\sigma) = \{\, (\sigma, i) \in \mathrm{Ob}(\mathcal{C}_\xi) \mid \text{$\sigma$ occurs in $\xi$ starting at position $i$} \,\}.
\end{equation}
This set records all positions where $\sigma$ appears in $\xi$. If $\sigma$ does not appear in $\xi$, then $S(\sigma) = \emptyset$.

Define $f_\xi(\sigma) = \# S(\sigma)$ as the number of occurrences of $\sigma$ in $\xi$. The function $f_\xi$ can be used to define a filtration on the set of sequences, which is useful for applying persistent homology and related topological methods.

To quantify the difference in occurrence frequency between two subsequences $\sigma_1$ and $\sigma_2$ within the same ambient sequence $\xi$, we define the \emph{frequency distance} as
\begin{equation}
d_{\mathrm{freq}}(\sigma_1, \sigma_2) = \left| f_\xi(\sigma_1) - f_\xi(\sigma_2) \right|.
\end{equation}

This distance captures the absolute difference in the number of times the two subsequences appear in $\xi$. A distance of zero indicates equal frequency, while larger values reflect greater disparities. The frequency distance is non-negative, symmetric, and satisfies the triangle inequality, thereby providing a simple yet informative way to compare patterns in sequential data.

\subsubsection*{Combined distance}

We can further enhance the analysis by combining both the content and position information of objects in the resolution category, thereby constructing a more comprehensive distance that captures both the structure of subsequences and their locations within the original sequence.

To this end, we define a \emph{combined distance} that integrates both the LCS distance and the position distance
\begin{equation}
d((\sigma_1, i_1), (\sigma_2, i_2)) = \alpha \cdot d_{\text{LCS}}(\sigma_1, \sigma_2) + \beta \cdot d_{\text{pos}}((\sigma_1, i_1), (\sigma_2, i_2)),
\end{equation}
where $\alpha, \beta \in \mathbb{R}$ are weights that control the relative importance of content and position differences. The values of $\alpha$ and $\beta$ can be chosen based on the specific application. For example, if the relative location of subsequences is crucial, one may assign a larger weight to $\beta$; conversely, if structural similarity is more important, a larger $\alpha$ may be preferred.

Although this section introduces several different distance measures, in our subsequent applications, we primarily adopt the position distance. The position distance better captures the fine-grained features of sequences of moderate length and is well suited for the tasks considered in this work. In comparison, the intersection distance is more appropriate for analyzing relatively short sequences, while the frequency distance is more efficient for sequences with lengths in the tens or even hundreds of millions. The use of the combined distance depends largely on the specific requirements of the task. Naturally, other distance measures can also be considered, as each possesses its own advantages and is suited to different application scenarios.

\subsection{Substructure complexes in the resolution category}

In Section \ref{section:resolution}, we introduce the resolution category to provide a categorical framework for describing sequences. This formalism offers a systematic approach to analyzing sequences from a categorical perspective. Building upon the resolution category, we further derive its geometric counterpart, namely the substructure complexes. These substructure complexes serve as a bridge to the topological study of sequences and facilitate the application of persistent homology.

Let $X$ be a non-empty finite set, and let $\xi$ be a fixed sequence with elements in $X$. Associated with $\xi$, we consider a resolution category $\mathcal{C}_{\xi}$, whose objects and morphisms encode the structural information of the sequence.

Assume that a distance function $d$ is defined on the set of objects of $\mathcal{C}_{\xi}$. Let $U \subseteq \operatorname{Ob}(\mathcal{C}_{\xi})$ be a finite collection of objects. For each real number $r \in \mathbb{R}$, we define a $\Delta$-complex $\mathcal{K}_r(U)$ as follows: for $p \geq 1$, a $p$-simplex in $\mathcal{K}_r(U)$ is a subset $\{\sigma_0, \dots, \sigma_p\} \subseteq U$ of $p+1$ objects such that $d(\sigma_i, \sigma_j) \leq r$ for all $0 \leq i, j \leq p$. And the vertex set of $\mathcal{K}_r(U)$ is $U/\sim$. Here, $\sim$ represents equivalence of objects at the same position, i.e., $(\sigma_{1},i)\sim (\sigma_{2},i)$ for any $\sigma_{1}$ and $\sigma_{2}$. This complex is called the \textit{substructure complex} associated with $U$.
Substructure complexes of a resolution category are also referred to as \textit{resolution complexes}.

Consider the case where $U = S(\sigma)$. In this setting, $\mathcal{K}_r(U)$ coincides with the Vietoris--Rips complex of the metric subspace $(U, d)$ at scale parameter $r$. We denote by \(\mathcal{K}^n_r(U)\) the set of all \(n\)-simplices in the complex \(\mathcal{K}_r(U)\).
As $r$ increases, this construction defines a filtration $\{ \mathcal{K}_r(U) \}_{r \in \mathbb{R}}$, which gives rise to persistent homology. The resulting persistence module captures the multi-scale topological features of the object collection $U$ within $\mathcal{C}_{\xi}$.

For any real numbers $a \leq b$, we obtain an embedding map of simplicial complexes $f^{a,b}: \mathcal{K}_a(U) \to \mathcal{K}_b(U)$. This induces a homomorphism on homology groups
\begin{equation}
f^{a,b}_{n}: H_n(\mathcal{K}_a(U)) \to H_n(\mathcal{K}_b(U)),
\end{equation}
where $H_n(-)$ denotes the singular homology with coefficients in a field $\mathbb{K}$. This gives rise to the concept of persistent homology.

\begin{definition}
  The $(a,b)$-persistent homology of $U$ in the resolution category $\mathcal{C}_{\xi}$ is defined as
  \begin{equation}
  H_{n}^{a,b}(U) \coloneqq \im \left( f^{a,b}_{n}: H_n(\mathcal{K}_a(U)) \to H_n(\mathcal{K}_b(U)) \right), \quad n \geq 0.
  \end{equation}
\end{definition}
In particular, when $a = b$, we have $H_n^{a,a}(U) = H_n(\mathcal{K}_a(U))$, which is precisely the homology of the complex $\mathcal{K}_a(U)$ at time $a$.

The family of groups $H_n^{a,b}(U)$ for $n \geq 0$ describes the persistence of topological features in the collection $U$ as the scale parameter $r$ evolves from $a$ to $b$. The concept of persistent homology captures how these features, such as connected components, loops, or higher-dimensional holes, appear and disappear as the filtration parameter changes. The 0-dimensional Betti number is used to characterize the number of connected components, the 1-dimensional Betti number represents the number of loops, and the 2-dimensional Betti number counts the number of voids (hollow cavities). We typically denote $\beta^{a,b}_{n}(U) = \mathrm{rank} \, H_n^{a,b}(U)$ to represent the number of topological features that persist from time $a$ to time $b$.

\begin{example}\label{substructure_complex}

Let $\xi = \text{ACTGTCAAGTTT}$ be a nucleic acid sequence, and let $U = \{C, G\}$ be the subset of nucleotides of interest. See Figure \ref{fig:ResPHworkflow}, the positions of $C$ and $G$ in $\xi$ are as follows: $C$ appears at positions $1$ and $5$, and $G$ appears at positions $3$ and $8$. Recall that the position distance between two objects $(\sigma_1, i_1)$ and $(\sigma_2, i_2)$ is defined as
\begin{equation}
d_{\mathrm{pos}}((\sigma_1, i_1), (\sigma_2, i_2)) = |i_1 - i_2|.
\end{equation}
Using this, we compute the position distances between all pairs of objects in $U = \{C, G\}$, where the objects correspond to nucleotides and their positions in the sequence. The resulting distance matrix is
\begin{equation}
\begin{pmatrix}
  & (C, 1) & (G, 3) & (C, 5) & (G, 8) \\
(C, 1) & 0 & 2 & 4 & 7 \\
(G, 3) & 2 & 0 & 2 & 5 \\
(C, 5) & 4 & 2 & 0 & 3 \\
(G, 8) & 7 & 5 & 3 & 0 \\
\end{pmatrix}
\end{equation}
This matrix represents the position distances between all pairs of nucleotides from $U$.

Next, we construct the substructure complex $\mathcal{K}_r(U)$ for a given parameter $r$. This complex consists of simplices formed by collections of objects, where any two objects $(\sigma_1, i_1)$ and $(\sigma_2, i_2)$ satisfy that their positional distance is less than or equal to $r$.
For simplicity, we denote the objects $(C, 1)$, $(G, 3)$, $(C, 5)$, and $(G, 8)$ as $v_1, v_2, v_3,$ and $v_4$, respectively. For the case $r = 2$, the substructure complex is
\begin{align*}
  \mathcal{K}^{0}_{2}(U) &= \{\{v_1\}, \{v_2\}, \{v_3\}, \{v_4\}\}, \\
  \mathcal{K}^{1}_{2}(U) &= \{\{v_1, v_2\}, \{v_2, v_3\}\}.
\end{align*}
This is a $1$-dimensional simplicial complex. Its homology is given by
\begin{equation}
H_n(\mathcal{K}_2(U)) =
\begin{cases}
\mathbb{K} \oplus \mathbb{K}, & \text{if } n = 0, \\
0, & \text{otherwise}.
\end{cases}
\end{equation}
For $r = 4$, the substructure complex becomes
\begin{align*}
  \mathcal{K}^{0}_{4}(U) &= \{\{v_1\}, \{v_2\}, \{v_3\}, \{v_4\}\}, \\
  \mathcal{K}^{1}_{4}(U) &= \{\{v_1, v_2\}, \{v_2, v_3\}, \{v_1, v_3\}, \{v_3, v_4\}\}, \\
  \mathcal{K}^{2}_{4}(U) &= \{\{v_1, v_2, v_3\}\}.
\end{align*}
This is a $2$-dimensional simplicial complex. The $(2,4)$-persistent homology of $U$ is
\begin{equation}
H_n^{2,4}(U) =
\begin{cases}
\mathbb{K}, & \text{if } n = 0, \\
0, & \text{otherwise}.
\end{cases}
\end{equation}
Thus, the corresponding persistent Betti numbers are $\beta^{2,4}_{0}(U)=1$ and $\beta^{2,4}_{n}(U)=0$ for $n\geq 1$.

Let $U = \{A, AG\}$ be the subset of nucleic acid patterns of interest. The positions of $A$ and $AG$ in $\xi$ are as follows: $A$ appears at positions $0$, $6$, and $7$, and $AG$ appears at position $7$. We associate these occurrences with labeled objects as follows: $(A,0)$, $(A,6)$, $(A,7)$, and $(AG,7)$. For simplicity, we denote these objects as $w_1 = (A,0)$, $w_2 = (A,6)$, $w_3 = (A,7)$, and $w_4 = (AG,7)$.

The resulting distance matrix is computed as
\begin{equation}
\begin{pmatrix}
  & (A, 0) & (A, 6) & (A, 7) & (AG, 7) \\
(A, 0) & 0 & 6 & 7 & 7 \\
(A, 6) & 6 & 0 & 1 & 1 \\
(A, 7) & 7 & 1 & 0 & 0 \\
(AG, 7) & 7 & 1 & 0 & 0 \\
\end{pmatrix}
\end{equation}
This matrix represents the position distances between all pairs of nucleic acid patterns from $U$.

Next, we construct the substructure complex $\mathcal{K}_r(U)$ for a given parameter $r$. For the case $r = 1$, the substructure complex is
\begin{align*}
  \mathcal{K}^{0}_{1}(U) &= \{\{w_1\}, \{w_2\}, \{w_3\}=\{w_4\}\}, \\
  \mathcal{K}^{1}_{1}(U) &= \{\{w_2, w_3\}, \{w_2, w_4\}, \{w_3, w_4\}\}, \\
  \mathcal{K}^{2}_{1}(U) &= \{\{w_2, w_3, w_4\}\}.
\end{align*}
This is a 2-dimensional simplicial complex with an isolated vertex $w_1$. Its homology is given by
\begin{equation}
H_n(\mathcal{K}_1(U)) =
\begin{cases}
\mathbb{K} \oplus \mathbb{K}, & \text{if } n = 0,\\
\mathbb{K} , & \text{if } n = 1,\\
0, & \text{otherwise}.
\end{cases}
\end{equation}
For $r = 6$, the substructure complex is
\begin{align*}
  \mathcal{K}^{0}_{6}(U) &= \{\{w_1\}, \{w_2\}, \{w_3\}= \{w_4\}\}, \\
  \mathcal{K}^{1}_{6}(U) &= \{\{w_1,w_2\},\{w_2, w_3\}, \{w_2, w_4\}, \{w_3, w_4\}\}, \\
  \mathcal{K}^{2}_{6}(U) &=  \{\{ w_2, w_3, w_4\}\}.
\end{align*}
The $(1,6)$-persistent homology of $U$ is given by
\begin{equation}
H_n^{1,6}(U) =
\begin{cases}
\mathbb{K}, & \text{if } n = 0, 1, \\
0, & \text{otherwise}.
\end{cases}
\end{equation}
Thus, the corresponding persistent Betti numbers are $\beta^{1,6}_{0}(U)=\beta^{1,6}_{1}(U)=1$ and $\beta^{1,6}_{n}(U)=0$ for $n\geq 2$.
\end{example}

For a given DNA sequence of length \(N\), the computational complexity for constructing a \(k\)-dimensional resolution complex is, in the worst case, \(O(kN)\). With the use of optimized algorithms, this complexity can be improved to \(O(N)\). For the entire pipeline, the overall computational complexity is typically $O(N \log N)$, dominated by Betti number computations. However, due to the special nature of the position distance, the use of advanced optimization techniques and iterative methods can reduce the complexity to nearly \(O(N)\).

\section{Results}

\begin{figure}[htbp!]
    \centering
    \includegraphics[width=\textwidth]{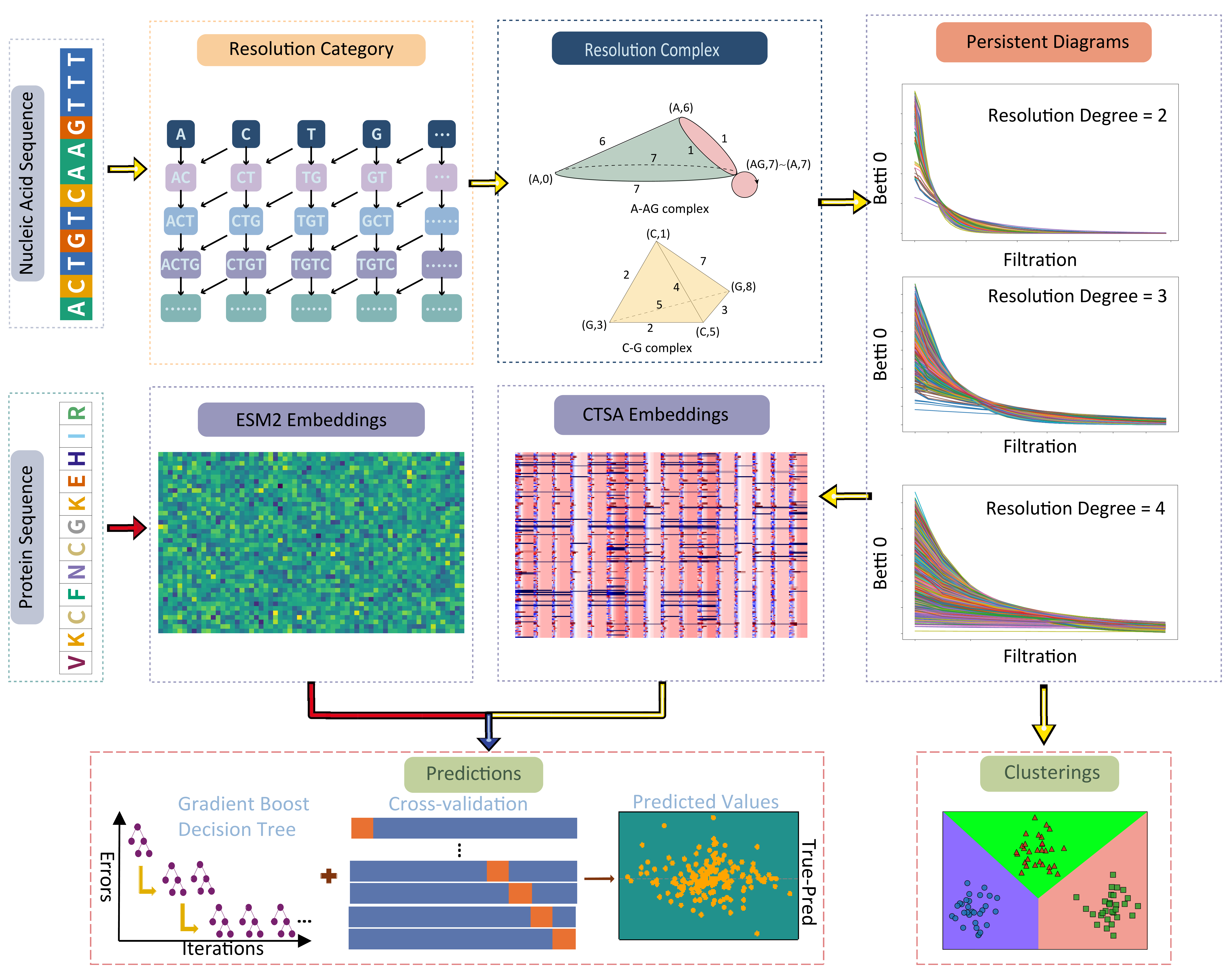}
    \caption{
    Overview of the CTSA framework. A nucleic acid sequence is first encoded into a resolution category that captures its multi-scale structure algebraically. Based on this, a resolution complex is constructed as a topological realization of the sequence. Persistent homology is then applied to extract multiscale topological features, forming the CTSA embeddings. Protein sequences are embedded using ESM2. For interaction prediction tasks, CTSA and ESM2 features are concatenated and used in supervised models. For clustering tasks, only CTSA features are used. Performance is evaluated through cross-validation and comparative variant analysis.
    }

    \label{fig:ResPHworkflow}
\end{figure}

Figure~\ref{fig:ResPHworkflow} illustrates the overall workflow of our proposed CTSA framework to extract topological representations from nucleic acid sequences. The foundation of this approach is the \emph{resolution category}, an algebraic structure that captures the multiscale organization of a sequence by formalizing relationships among its subsequences.

From this algebraic structure, we construct the \emph{resolution complex}, a $\Delta$-complex that serves as the topological realization of the resolution category and, by extension, of the sequence itself. Persistent homology is then applied to this complex, producing topological features (e.g. Betti numbers) that describe the sequence's intricate topological structures across multiple resolution scales.
For protein-nucleic acid interaction tasks, these features are integrated with transformer-based embeddings of protein sequences (e.g., ESM2). The resulting feature representations are used in downstream predictive and clustering analyses.

This topological framework enables principled multiscale analysis of sequence structure and provides a unified pipeline for quantitative and qualitative biological inference.

\subsection{CTSA for predictions of protein-nucleic acid binding affinities}

	Protein-nucleic acid interactions represent a fundamental class of biomolecular bindings essential for numerous biological processes. These interactions play a vital role in controlling the flow of genetic information by enabling signal transduction, translation, transcription, replication, and repair \cite{luscombe2000overview}.  Moreover, this interaction between protein and nucleic acids is central to regulating the structural integrity of gene expression, thus ensuring well-regulated cellular activity. Proteins mediate molecular-level mechanisms by recognizing and binding to specific DNA or RNA sequences through mechanisms such as electrostatic attraction, hydrogen bonding, physicochemical forces, and structural complementarity~\cite{steitz1990structural}. Disruption in such binding events can contribute to the development of various diseases, such as cancer, neurodegenerative conditions, and autoimmune disorders. As a result, understanding protein-nucleic acid binding is crucial not only for decoding biomolecular functions but also for paving the way for new therapies and drug discovery.

	Sequence-based virtual screening is a computational approach that relies solely on the primary sequences of biomolecules, such as proteins or nucleic acids, to predict their binding affinities or interactions, without relying on 3D structural data. In this study, we employ Shen et al's dataset~\cite{shen2023svsbi} which includes 186 protein-nucleic acid complexes constructed from PDBbind v2020. We then implemented CTSA framework to analyze protein-nucleic acid binding affinity by extracting transformer-based sequence embeddings and resolution-based topological sequence features.

    For proteins, we employ ESM2~\cite{lin2022language}, a state-of-the-art transformer-based language model with approximately 3 billion parameters and trained on millions of protein sequences with 50\% sequence identity. We generate a total of 2560 high-dimensional protein embeddings, each capturing evolutionary patterns and deep relationships across 36 transformer layers.

    For nucleic acids, we validate CTSA in DNA sequences to assess its capability to reveal essential interaction patterns relevant to protein binding. CTSA provides a topological realization of DNA by introducing resolution categories that reflect the intricate connections of the sequence within its k-mer spaces.  Persistent homology is then applied across these categories to extract topological features, namely, persistent Betti numbers, which are essential for CTSA embeddings to capture both local and global structure of sequence. The resulting CTSA feature vectors have a dimension of 800, which is comparable to the 768-dimensional DNABERT features used in the related SVSBI study.

    By concatenating the ESM2-based protein embeddings and CTSA-based DNA embeddings, we construct unified feature vectors for each protein-nucleic acid pair. These vectors are then used in downstream machine learning models to predict binding affinity.

\begin{figure}[htbp]
\centering
\includegraphics[width=\textwidth]{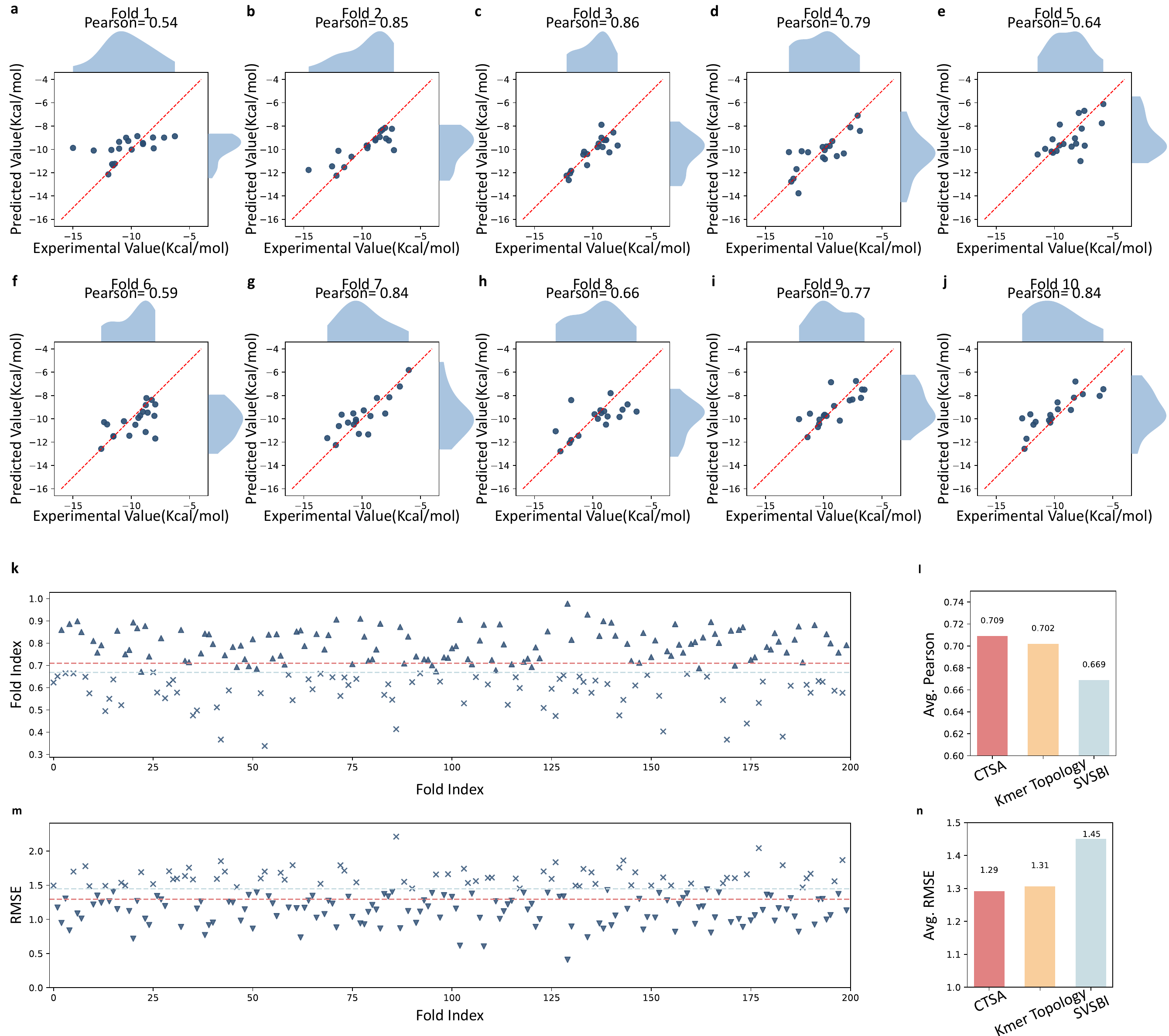}
\caption{
Overview and evaluation of the CTSA protein-nucleic acid binding affinity prediction framework.
\textbf{a}-\textbf{j} Scatter plots of predicted versus experimental binding affinities for each fold in one representative round of 10-fold cross-validation, illustrating the predictive accuracy of the model.
\textbf{k} Fold-wise Pearson correlation results across 20 rounds of cross-validation for CTSA.
\textbf{m} Fold-wise RMSE across 20 rounds of cross-validation for CTSA.
\textbf{l} and \textbf{n} Comparison of average Pearson correlation and RMSE of CTSA and baseline methods, respectively, over all 20 rounds.
These plots together demonstrate the robustness and superior performance of CTSA in modeling sequence-based interactions.
}\label{fig:PNbinding}
\end{figure}

    To evaluate the predictive power of our CTSA framework, we conducted 20 rounds of 10-fold cross-validation on the benchmark dataset.   Figures~\ref{fig:PNbinding}\textbf{a}-\textbf{j} illustrate the predicted versus experimental binding affinities generated from our CTSA model. The model achieves strong correlations across folds, with Pearson coefficients ranging from 0.54 to 0.86. While some variability is observed, an expected outcome given the small dataset size, the majority of folds exceed 0.70, indicating that the model reliably captures the underlying interaction patterns in protein-nucleic acid complexes.

    To comprehensively assess model robustness, we aggregate results from all 20 rounds. Figure~\ref{fig:PNbinding}\textbf{k} presents the fold-wise Pearson correlations across repetitions, showing that CTSA consistently outperforms SVSBI and the k-mer topology baseline. Similarly, fold-wise RMSE comparisons are shown in Figure~\ref{fig:PNbinding}\textbf{m}, where CTSA exhibits lower prediction errors in most folds.

    The overall average performance across all runs is summarized in Figure~\ref{fig:PNbinding}\textbf{l} for Pearson correlation and Figure~\ref{fig:PNbinding}\textbf{n} for RMSE. CTSA achieves a mean Pearson correlation of 0.709 and an RMSE of 1.29 kcal/mol, outperforming SVSBI~\cite{shen2023svsbi} (0.669 / 1.45 kcal/mol) and the k-mer topology~\cite{hozumi2024revealing} baseline (0.702 / 1.31 kcal/mol). These results validate the effectiveness and stability of CTSA in capturing interaction-relevant features from DNA sequences and enabling accurate protein-nucleic acid binding affinity prediction.

\subsection*{Phylogenetic analysis of SARS-CoV-2 Variants}

To further validate the effectiveness of CTSA in phylogenetics, we apply our method to a clustering task involving SARS-CoV-2 genomes. Classifying SARS-CoV-2 variants poses a significant challenge for alignment-free methods, as the distinguishing mutations between variants often involve only a few nucleotide differences. Specifically, we analyze 44 complete SARS-CoV-2 sequences obtained from GISAID, labeled according to their known variants: Alpha, Beta, Gamma, Delta, Lambda, Mu, GH/490R, and Omicron.

Analyzing variants within a single species is particularly challenging for alignment-free methods, as the differences are often limited to a few key nucleotide mutations. These mutations, however, are known to significantly affect infectivity and transmission. We assess the ability of CTSA to group these sequences accurately into their corresponding variant clades and compare its performance with five state-of-the-art alignment-free approaches: natural vector method (NVM)~\cite{deng2011novel}, Jensen-Shannon divergence (FFP-JS)~\cite{jun2010whole,sims2009alignment}, Kullback-Leibler divergence(FFP-KL)~\cite{vinga2003alignment}, Markov K-String~\cite{qi2004whole}, and Fourier Power Spectrum (FPS)~\cite{hoang2015new}.

\begin{figure}[htbp!]
    \centering
    \includegraphics[width=\textwidth]{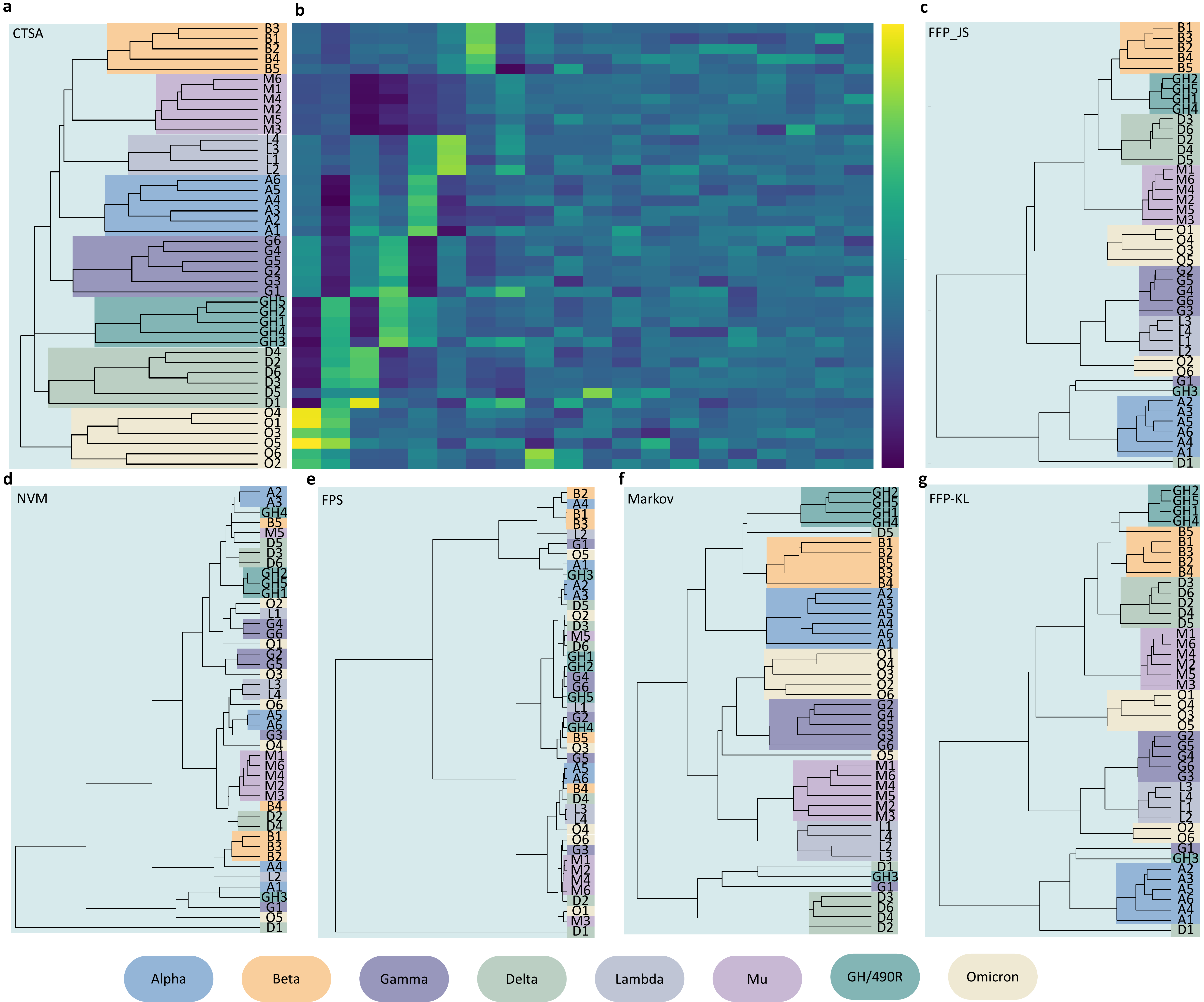}
\caption{
Comparison of alignment-free methods for phylogenetic analysis and feature space clustering of 44 complete SARS-CoV-2 genomes. The sequences are categorized by known variants: Alpha, Beta, Gamma, Delta, Lambda, Mu, GH/490R, and Omicron.
\textbf{a} Phylogenetic tree generated by CTSA, showing hierarchical relationships among sequences based on topological features.
\textbf{b} PCA visualization of CTSA feature embeddings used in \textbf{a}, illustrating their separability.
\textbf{c}-\textbf{g} Phylogenetic trees produced by five baseline methods-NVM, FFP-JS, FFP-KL, Markov K-String (Markov), and Fourier Power Spectrum (FPS).
}
\label{fig:sarsClustering}
\end{figure}

Figure~\ref{fig:sarsClustering}\textbf{a}  shows that CTSA successfully clusters all 44 sequences into their correct variant clades. To further examine the separability of the learned features, Figure~\ref{fig:sarsClustering}\textbf{b}  presents a PCA projection~\cite{mackiewicz1993principal} of the CTSA feature representations used to generate the phylogenetic tree in Figure~\ref{fig:sarsClustering}\textbf{a}.  The visualization reveals that sequences from the same variant cluster closely together in feature space, reinforcing that CTSA captures discriminative sequence characteristics relevant to variant classification.

In contrast, the other methods exhibit varying degrees of misclassification: NVM (Figure~\ref{fig:sarsClustering}\textbf{d}) forms only partial clusters; FFP-JS and FFP-KL (Figure~\ref{fig:sarsClustering}\textbf{c} and Figure~\ref{fig:sarsClustering}\textbf{e}) misclassify three sequences each and split Omicron into two clades; the Markov method (Figure~\ref{fig:sarsClustering}\textbf{f}) misclassifies five sequences; and FPS (Figure~\ref{fig:sarsClustering}\textbf{e})  fails to form meaningful clusters altogether.

Note that persistent homology-based  k-mer topology does not offer completely accurate clustering of SARS-CoV-2 variants\cite{hozumi2024revealing}. It has errors with the Delta variant. The k-mer topology model needs the help of persistent Laplacians to achieve 100\% accuracy on this dataset. Whereas, the proposed CTSA  achieves completely accurate clustering by using persistent homology. Therefore, CTSA is more accurate than k-mer topology.

These results demonstrate the superior clustering performance of CTSA, especially in detecting subtle sequence differences critical for variant classification, highlighting its potential as a robust alignment-free tool in viral genome analysis.

\section{Concluding remarks}

Sequence data inherently exhibit complex and often ambiguous structural characteristics, which are challenging to uncover using conventional methods. Topological tools, known for their robustness in handling high-dimensional and intricate data, offer unique advantages in such contexts. Building upon this foundation, we propose a foundational mathematical framework for analyzing sequence data from a topological perspective. More specifically, we model a sequence using the language of category theory. Given a sequence, we construct a resolution category that encodes the hierarchical organization and structural information of the sequence. This categorical representation serves as the basis for building substructure complexes, whose persistent homology captures the topological features of the sequence across multiple scales.

In this work, the proposed category-based topological sequence analysis (CTSA) serves as a novel framework for analyzing sequence data through a topological lens. CTSA represents a conceptual shift in the design of alignment-free methods, moving beyond frequency-based or embedding-based representations to incorporate structured mathematical formalisms grounded in the topology of sequences. By leveraging categorical and homological tools, CTSA offers a generalizable and principled approach for encoding and analyzing the complex, hierarchical architecture of biological sequences without relying on sequence alignment. This framework captures topological signatures that are often elusive to traditional techniques and provides a multiscale view of discrete sequence structure. One key insight from our results is that the resolution of local-to-global sequence structure, captured through the filtration of substructure complexes, is critical for both predictive modeling and comparative tasks. The fact that CTSA achieves strong performance in tasks with fundamentally different goals and data characteristics suggests that the topological information it captures is broadly informative, not task-specific. Indeed, CTSA is a generalization of k-mer topology \cite{hozumi2024revealing}  and is designed to be more accurate than k-mer topology.

From a methodological perspective, CTSA demonstrates the potential of applying resolution hierarchies and categorical abstractions to model discrete biological structures. This framework invites future exploration into its adaptation for sequence-structure relationships, non-coding regulatory region analysis, and as a structured, multiscale representation for large language models in genomics-domains where effectively capturing hierarchical biological context remains a major challenge. Although our study primarily focused on biological sequences, the methodology is modality-agnostic and could, in principle, be extended to non-biological sequential data.

\section*{Data and Code Availability}
The data and source code obtained in this work are publicly available in the Github repository: \hyperlink{https://github.com/WeilabMSU/ResPH}{https://github.com/WeilabMSU/ResPH}

\section*{Author Contributions}
Jian Liu wrote the first draft, developed the theoretical framework and methods, and revised the manuscript. Shen Li wrote the first draft, designed the algorithms, created the figures, and participated in revision. Mushal Zia contributed to the theoretical analysis, data collection, and algorithm development. Guo-Wei Wei conceived the main ideas, supervised the project, revised the manuscript, and secured funding.

\section*{Funding Information}
This work was supported in part by NIH grants R01GM126189, R01AI164266, and R35GM148196, National Science Foundation grants DMS2052983, DMS-1761320, and IIS-1900473, NASA  grant 80NSSC21M0023,   Michigan State University Research Foundation, and  Bristol-Myers Squibb  65109. Jian was also supported by Natural Science Foundation of China (NSFC Grant No. 12401080) and Scientific Research Foundation of Chongqing University of Technology.

\section*{Conflict of Interest}
The authors declare no competing interests.

\section*{Statement of Usage of Artificial Intelligence}
Artificial intelligence tools or technologies were not used in this project except for limited English grammar corrections.

\bibliographystyle{plain}  
\bibliography{Reference}

\end{document}